\documentclass[journal]{IEEEtran}

\usepackage{cite}
\usepackage{amsmath,amssymb,amsfonts}
\usepackage{algorithmic}
\usepackage{graphicx}
\usepackage{textcomp}
\usepackage{wrapfig}
\usepackage[utf8]{inputenc}
\usepackage{graphicx}
\usepackage{cite}
\usepackage{indentfirst}
\usepackage[justification=centering]{caption}
\usepackage{amsmath}
\usepackage{lscape}
\usepackage{gensymb}
\usepackage{algorithm}
\usepackage{algorithmic}
\usepackage{mathtools}
\usepackage{amsfonts}
\usepackage{comment}
\usepackage{bm}
\usepackage{multirow}
\usepackage{marvosym}
\usepackage[dvipsnames]{xcolor}
\usepackage{multirow}
\usepackage{array}
\usepackage{makecell}
\usepackage[colorlinks=true, linkcolor=black, citecolor=black, urlcolor=black]{hyperref}
\def\BibTeX{{\rm B\kern-.05em{\sc i\kern-.025em b}\kern-.08em
    T\kern-.1667em\lower.7ex\hbox{E}\kern-.125emX}}

\begin{document}

\title{PCA-Guided Autoencoding for Structured Dimensionality Reduction in Active Infrared Thermography}

\author{Mohammed Salah,
        Numan Saeed,
        Davor Svetinovic,
        Stefano Sfarra,
        Mohammed Omar,
        and Yusra Abdulrahman
\thanks{This work was supported by Khalifa University of Science and Technology under Award 8474000660. The work was also supported by the Advanced Research and Innovation Center (ARIC), which is jointly funded by Aerospace Holding Company LLC, a wholly-owned subsidiary of Mubadala Investment Company PJSC and Khalifa University for Science and Technology.}
\thanks{M. Salah and Y. Abdulrahman are with the Department of Aerospace Engineering, Khalifa University, Abu Dhabi, UAE. Y. Abdulrahman is also with the Advanced Research and Innovation Center (ARIC), Khalifa University, Abu Dhabi, UAE.

Numan Saeed is with BioMedIA Lab in Mohamed Bin Zayed University of Artificial Intelligence (MBZUAI), Abu Dhabi, UAE.

D. Svetinovic is with the Department of Electrical Engineering and Computer Science, Khalifa University of Science and Technology, Abu Dhabi, UAE. D. Svetinovic is also with the Department of Information Systems and Operations Research, Vienna University of Economics and Business, 1090 Vienna, Austria.

S. Sfarra is with the Department of Industrial and Information Engineering and Economics (DIIIE), University of L’Aquila, L'Aquila I-67100, Italy

M. Omar is with the Department of Industrial and Systems Engineering, Khalifa University, Abu Dhabi, UAE.

Yusra Abdulrahman is the corresponding author (email: yusra.abdulrahman@ku.ac.ae).}}

\IEEEaftertitletext{\vspace{-2\baselineskip}}

\maketitle

\begin{abstract}
Active Infrared thermography (AIRT) is a widely adopted non-destructive testing (NDT) technique for detecting subsurface anomalies in industrial components. Due to the high dimensionality of AIRT data, current approaches employ non-linear autoencoders (AEs) for dimensionality reduction. However, the latent space learned by AIRT AEs lacks structure, limiting their effectiveness in downstream defect characterization tasks. To address this limitation, this paper proposes a principal component analysis guided (PCA-guided) autoencoding framework for structured dimensionality reduction to capture intricate, non-linear features in thermographic signals while enforcing a structured latent space. A novel loss function, PCA distillation loss, is introduced to guide AIRT AEs to align the latent representation with structured PCA components, while capturing the intricate, non-linear patterns in thermographic signals. To evaluate the utility of the learned, structured latent space, we propose a neural network–based evaluation metric that assesses its suitability for defect characterization. Experimental results show that the proposed PCA-guided AE outperforms state-of-the-art dimensionality reduction methods on PVC, CFRP, and PLA samples in terms of contrast, signal-to-noise ratio (SNR), and neural network-based metrics.
\end{abstract}


\IEEEpeerreviewmaketitle

\section*{Code:}
\begin{center}
\url{https://github.com/mohammedsalah98/pca_guided_ae}
\end{center}

\section{Introduction}
In response to the increasing demand for reliable quality assurance across modern industries, non-destructive testing (NDT) techniques have emerged for predictive safety monitoring in various production lines, such as the aerospace \cite{ndt_review, 3d_cnn} automotive \cite{automotive_1,automotive_2}, and construction \cite{construction_ndt}. Carbon Fiber-Reinforced Polymers (CFRPs), Polylactic Acid (PLA), and Polyvinyl Chloride (PVC) materials have gained significant use in the aforementioned industries due to their lightweight nature, cost-effectiveness, and versatility in manufacturing. Nevertheless, subsurface anomalies can exist in components composed of these materials, which significantly reduce their lifespan \cite{ndt_survey, tip_ndt}. As a result, a diverse set of NDT techniques are continuously investigated and explored for improved quality control, including radiography inspection, ultrasonic testing, and infrared thermography (IRT) \cite{ndt_review}. 

IRT is currently attracting significant attention and is valued for its ability to rapidly inspect large areas and detect subsurface defects without requiring physical contact \cite{yusra_taguchi}. These advantages drove the exploration of the potential of IRT in inspecting aircraft components \cite{thermosense_artwork, cfrp_deep, 3d_cnn, yusra_4}, construction pipelines \cite{construction_ndt, construction_uav, uav_solar, drone_sites}, and artworks \cite{autoencoder, artwork}. Among IRT methodologies, Active IRT (AIRT) is heavily explored for inspecting subsurface anomalies in critical industrial components. In addition, AIRT is currently witnessing the adoption of artificial intelligence (AI) methodologies for automating and enhancing inspection accuracy. Accordingly, applications of AI-based AIRT have been developed for defect classification \cite{attention_spatiotemporal, irt_depth, tip_imaging}, defect segmentation \cite{pt_dataset, attention_unet}, and structural health monitoring \cite{construction_uav}.

In all of the aforementioned applications, data compression techniques are employed to reduce the high-dimensionality of AIRT data. The most prevalent state-of-the-art AIRT approaches rely on principal component analysis (PCA) to compress thermographic sequences before feeding them into AI models. PCA is shown to enhance the visibility of subsurface defects for precise defect analysis \cite{pca_2, sparse_kernel_pca, pca_tii}. Despite its effectiveness for many applications, PCA remains a linear method and struggles to capture the nonlinear patterns in thermographic sequences. These non-linear features are critical for detecting subtle anomalies and for structural health assessment. To address this drawback, non-linear AIRT autoencoders (AEs) have been proposed to capture the non-linear temporal dynamics of thermographic signals \cite{autoencoder, artwork, 1d_cnn}. In these models, the compressed latent space is used as a compact representation for AI-based downstream defect analysis. Yet, a key limitation remains: \textit{AIRT AEs are non-discriminative, and the learned latent space lacks structure and consistency, reducing its effectiveness in AI-based downstream defect characterization tasks.} In general, neural networks and learning-based approaches require input representations that are structurally consistent for efficient learning. To address the drawback of conventional AIRT AEs, this paper proposes a PCA-Guided AE that combines the non-linear modeling capacity of AEs with the structured interpretability of PCA. The contributions of the paper are the following:

\begin{enumerate}
    \item A PCA-Guided AE is proposed to learn a compact and structured latent space from AIRT sequences for improved interpretability and discriminability.
    \item The PCA distillation loss is introduced to align the AE latent space with PCA components, while fully modelling the non-linearities of thermographic signals.
    \item A neural network–based evaluation metric is introduced to assess the utility of the proposed AE latent representation for downstream AI-based defect analysis.
    \item The proposed framework is tested on CFRP, PLA, and PVC specimens. Results show that the proposed framework outperforms the state-of-the-art in terms of contrast, signal-to-noise ratio (SNR), and neural network–based evaluation metrics.
    \item The source code of the PCA-Guided AE is released for the community to facilitate future benchmarking and research.
\end{enumerate}

\subsection{Related Work}
AIRT has been a critical NDT technique for detecting subsurface anomalies in a wide range of industrial components. This NDT technique is currently witnessing the adoption of AI methodologies to increase its reliability and defect detection accuracy. Thus, the applications of AI-driven AIRT are rapidly expanding in the aerospace \cite{cfrp_deep}, construction \cite{construction_ndt, thermosense_concrete}, and automotive \cite{automotive_1, automotive_2, automotive_3} industries. As AIRT applications are rapidly growing, a series of AIRT deep neural networks have been proposed. To illustrate, the Faster R-CNN and Yolov5 networks have been fine-tuned for detecting hidden defects in CFRP structures \cite{flexible_framework, irt_depth}. More advanced network architectures have also been proposed, such as ConvLSTM \cite{cnnlstm}, tailored for IRT data to further enhance defect detection accuracies compared to the previous approaches.

In addition to AIRT learning-based defect detection, AIRT learning-based defect segmentation has been investigated, with a range of segmentation network architectures introduced. Fang et al. \cite{unet_study} studied state-of-the-art segmentation networks, such as U-Net and ResNet, for subsurface defect segmentation in AIRT. On the other hand, the attention U-Net was adopted for improved defect segmentation compared to its traditional U-Net counterpart \cite{attention_unet}. Nevertheless, these networks learn filters that extract spatial features, disregarding the temporal aspect in thermographic sequences. Hence, 3D-CNNs were introduced to simultaneously extract spatial and temporal features from AIRT data for enhanced segmentation of subsurface defects \cite{3d_cnn}. Still, the aforementioned network is fully convolutional and tends to underperform due to the limited receptive field of CNNs. To overcome this challenge, attention mechanisms are employed in AIRT networks, capturing the long-range dependencies in AIRT data \cite{thermosense_attention}. Other methods also incorporate fusion strategies to harness the advantages of different thermographic representations for defect segmentation \cite{tip_fusion, pt_fusion}.

The aforementioned approaches rely on AIRT data compression techniques to generate thermographic representations as inputs to the deep neural networks. The most prominent AIRT dimensionality reduction techniques include thermographic signal reconstruction (TSR) \cite{tsr, tsr_2, tsr_3}, pulse phase thermography (PPT) \cite{ppt_1, ppt_2}, principal component analysis \cite{pct, pca_2, pca_tii, sparse_kernel_pca}. PCA is valued for its capability to enhance the visibility of barely detectable subsurface defects. However, as a linear method, PCA struggles to capture the non-linear features present in thermographic sequences. Thus, the denoising deep autoencoder thermography (DAT) and 2D deep convolutional autoencoder thermography (2D-DCAT) have been proposed for learning-based dimensionality reduction \cite{dat, autoencoder}. Still, the 2D-DCAT demands extensive computational resources during training. On the other hand, Zhang et al. \cite{1d_cnn} proposed a 1D convolutional autoencoder (1D-DCAE-AIRT) that is easier to train without sacrificing accuracy. However, the 1D-DCAE-AIRT architecture is fully convolutional, and its performance is constrained by the limited receptive field of the convolutional layers, failing to capture long-range dependencies in the data. More importantly, all of the previous learning-based dimensionality reduction techniques are non-discriminative and generate latent representations that lack consistency and structure. This is critical for utilizing such features as inputs to neural networks for downstream defect analysis tasks. This work addresses these challenges by proposing PCA-Guided AE that leverages the ability of autoencoders to model the non-linear temporal behavior of AIRT data while introducing structural guidance through alignment with PCA principal components. This hybrid approach enables the extraction of compact, interpretable, and task-relevant representations, enhancing their utility for learning-based defect characterization.

\subsection{Structure of The Article}
The rest of the article is structured as follows. Section \ref{sec:preliminaries} provides preliminaries for the paper. Section \ref{sec:methodology} outlines the methodology of the PCA-Guided autoencoder. Section \ref{sec:exps} presents experimental validations of the proposed framework. Finally, section \ref{sec:conc} presents conclusions, findings, and future work.

\section{Preliminaries} \label{sec:preliminaries}
\subsection{Active Infrared Thermography}

\begin{figure}[t]
\center
\includegraphics[keepaspectratio=true,scale=0.45, width=0.9\linewidth]{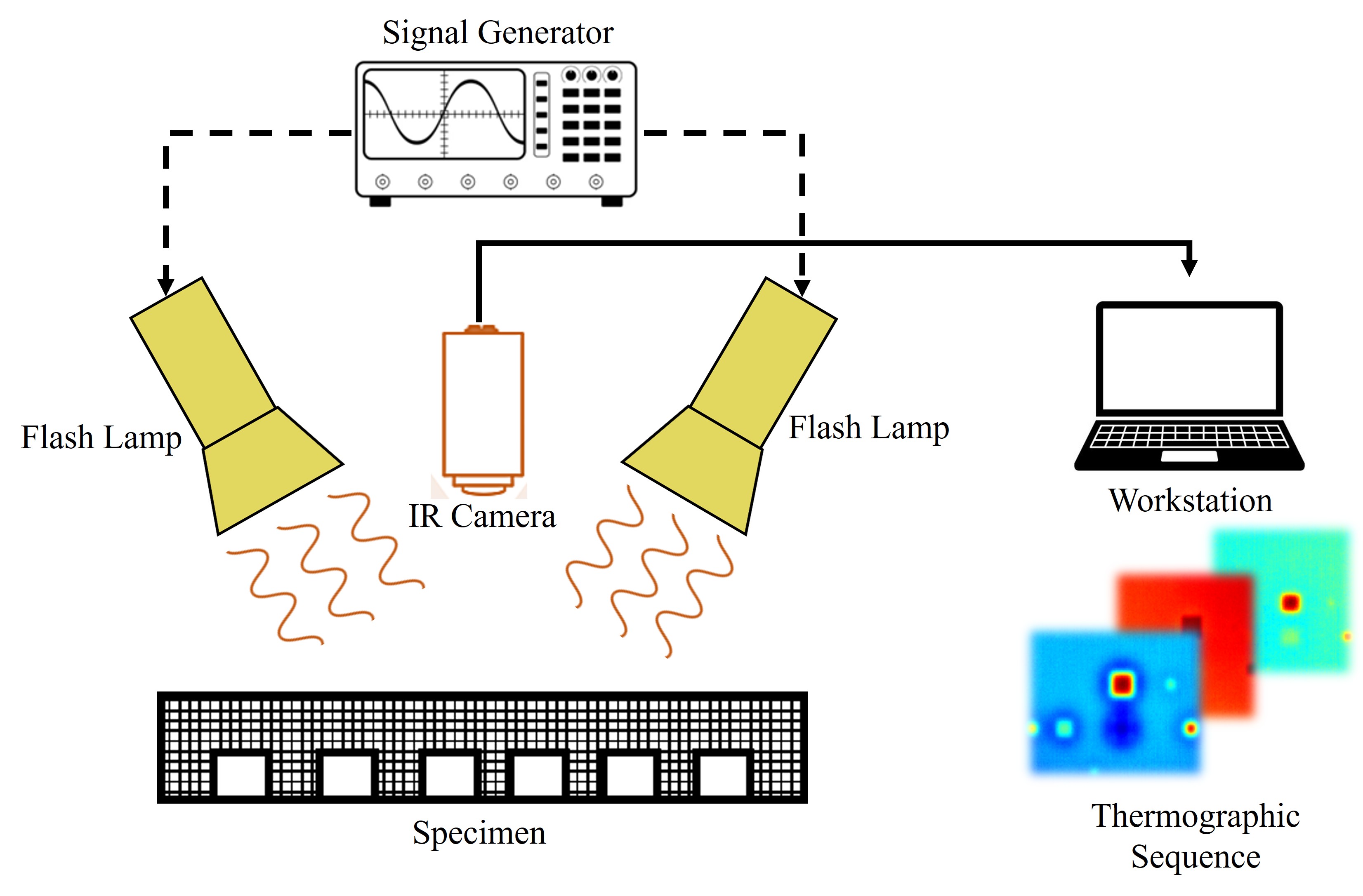}
\caption{Typical AIRT setup, where defective areas trap heat dissipation, generating abnormal thermal profiles detected by an IR camera.}
\label{fig:airt_setup}
\end{figure}

A standard AIRT configuration is illustrated in Fig. \ref{fig:airt_setup}. In this setup, a high-power halogen lamp delivers a controlled thermal excitation to the surface of the specimen \cite{ndt_review}. For defect-free regions, an IR camera records uniform thermal responses across all pixels. In the presence of defects, heat is trapped, resulting in abnormal thermal anomalies on the surface. These deviations are captured by the IR camera and used for subsequent analysis. Note that the specific nature of the thermal response depends on the type of excitation applied; for instance, pulse thermography employs a heat pulse, while lock-in thermography uses periodic heating for phase-based analysis. In both excitation paradigms, a 3D matrix, $\mathbf{S} = \{ I_{k} \}_{1}^{N_{t}}$, of shape $(N_{t}, N_{y}, N_{x})$ is generated, where $I_{k}$ is a thermogram timestamped at $k = 1, 2, \dots, N_{t}$, $N_{y}$ is the image height, and $N_{x}$ is its width. For subsequent defect analysis, $\mathbf{S}$ is reshaped to $(N_{t}, N_{y} \times N_{x})$ by a raster-like operation and standardized by

\begin{equation}
    \mathbf{\hat{S}} = \frac{\mathbf{S} - \mu_{k}}{\sigma_{k}},
\end{equation}

\noindent where,
\begin{equation}
    \mu_{k} = \frac{1}{N_{t}} \sum_{k=1}^{N_{t}} S^{(k)},
\end{equation}

\begin{equation}
    \sigma_{k}^{2} = \frac{1}{N_{t} - 1} \sum_{k=1}^{N_{t}}(S^{(k)} - \mu_{k})^{2},
\end{equation}

\noindent and $\mathbf{\hat{S}} = \{ S^{(n)} \}_{1}^{N_{x}\times N_{y}}$ is a matrix consisting of the standardized pixel-wise thermal response.

\subsection{Autoencoder Networks}
Prior to analyzing the standardized pixel responses, $\mathbf{\hat{S}}$, AIRT approaches perform dimensionality reduction, such as PCA, due to the high-dimensionality of $\mathbf{\hat{S}}$. PCA linearly projects high-dimensional data onto a lower-dimensional subspace defined by orthogonal directions of maximum variance and generates PCA images with amplified defect-relevant features. To generate the PCA images, singular value decomposition (SVD) is applied by

\begin{equation}
    \mathbf{\hat{S}} = \mathcal{U}\Gamma V^{T},
\end{equation}

\noindent where $\mathcal{U}$ contains the empirical orthogonal functions that capture spatial variations in the thermal responses, $\Gamma$ is a diagonal matrix of singular values arranged in descending order, and $V$ holds the corresponding principal component vectors. Finally, the $k^{th}$ principal component image, $P_k$, is obtained by projecting $\mathbf{\hat{S}}$ onto the $k^{th}$ eigenvector $v_k$ as

\begin{figure}[t]
\center
\includegraphics[keepaspectratio=true,scale=0.45, width=\linewidth]{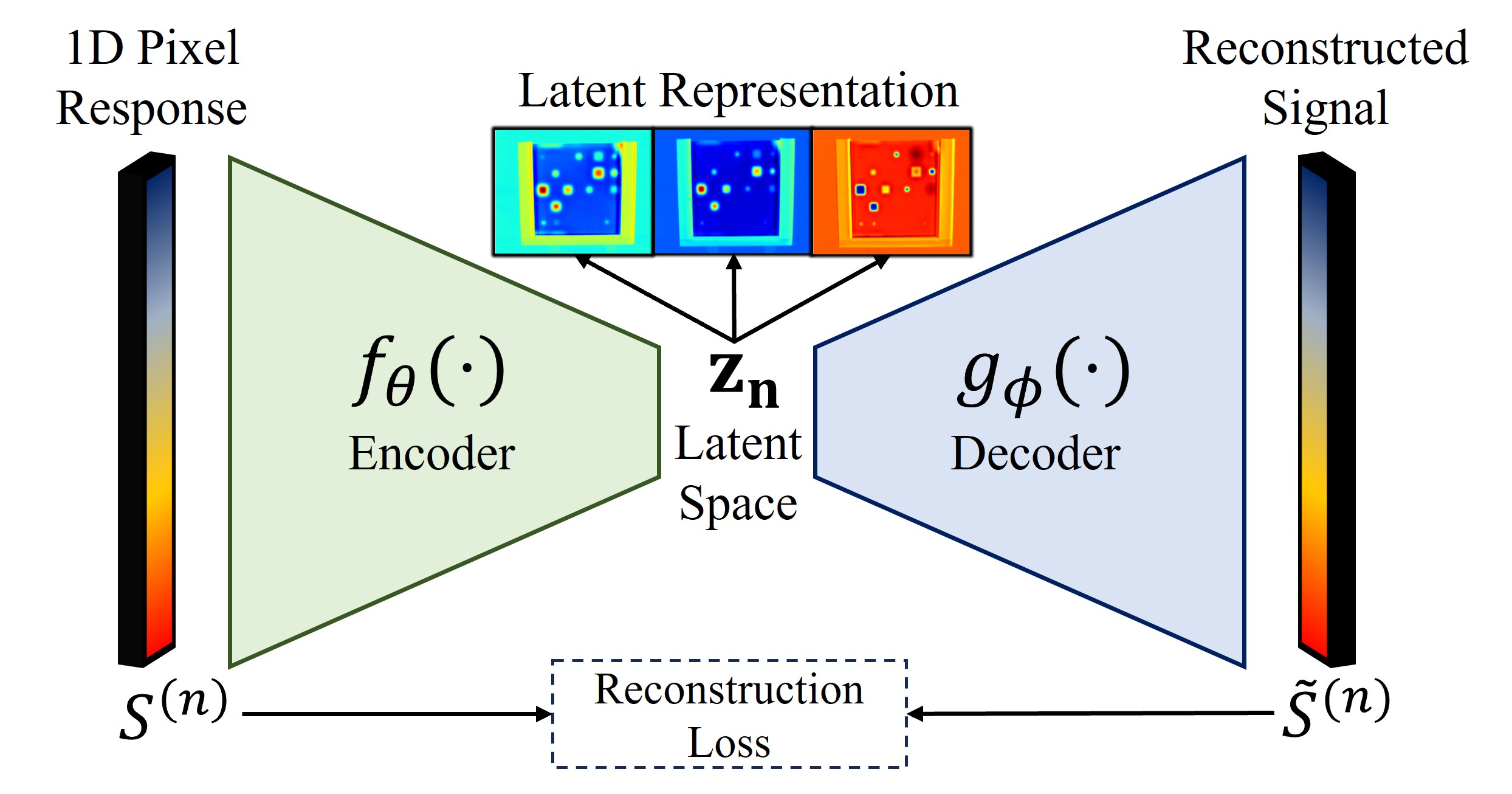}
\caption{Architecture of a standard AIRT autoencoder.}
\label{fig:ae_arch}
\end{figure}

\begin{figure}[b]
    \centering
    \includegraphics[keepaspectratio=true,scale=0.45, width=0.9\linewidth]{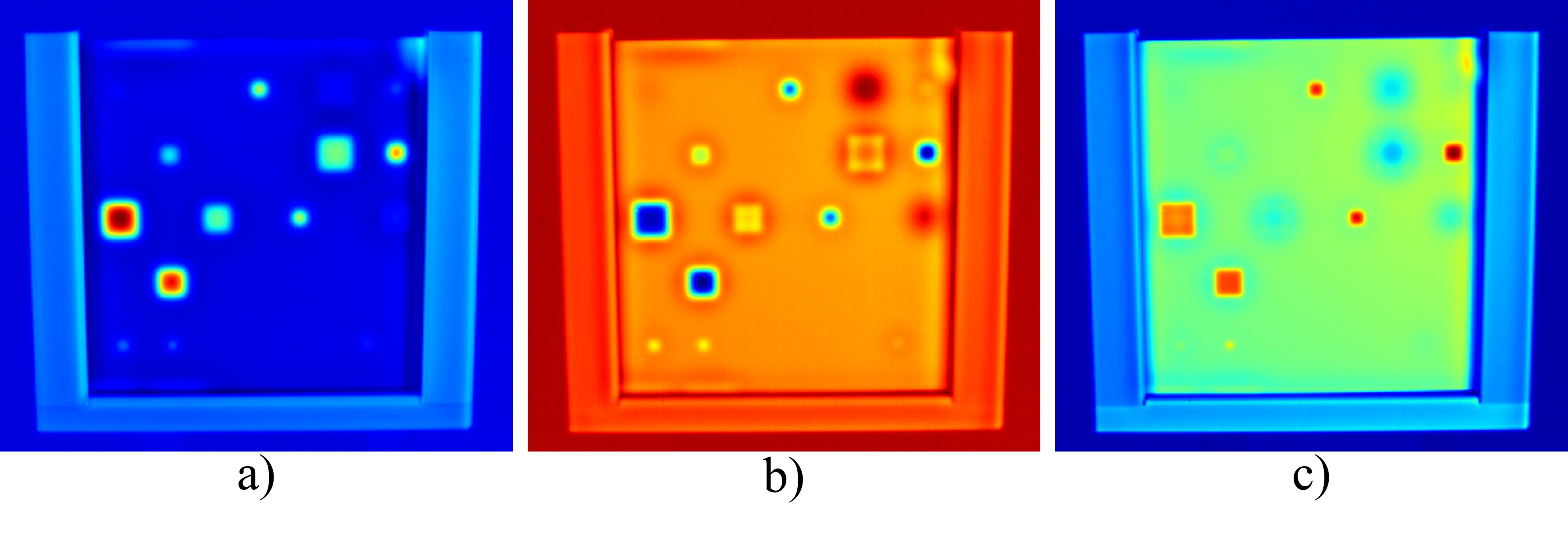}
    \caption{Latent representations extracted from a conventional AIRT autoencoder across different training runs, a)-c). The observed inconsistencies in the latent space highlight the lack of structure in the obtained representation.}
    \label{fig:latents}
\end{figure}

\begin{figure*}[!ht]
    \centering
    \includegraphics[keepaspectratio=true,scale=0.45, width=0.9\linewidth]{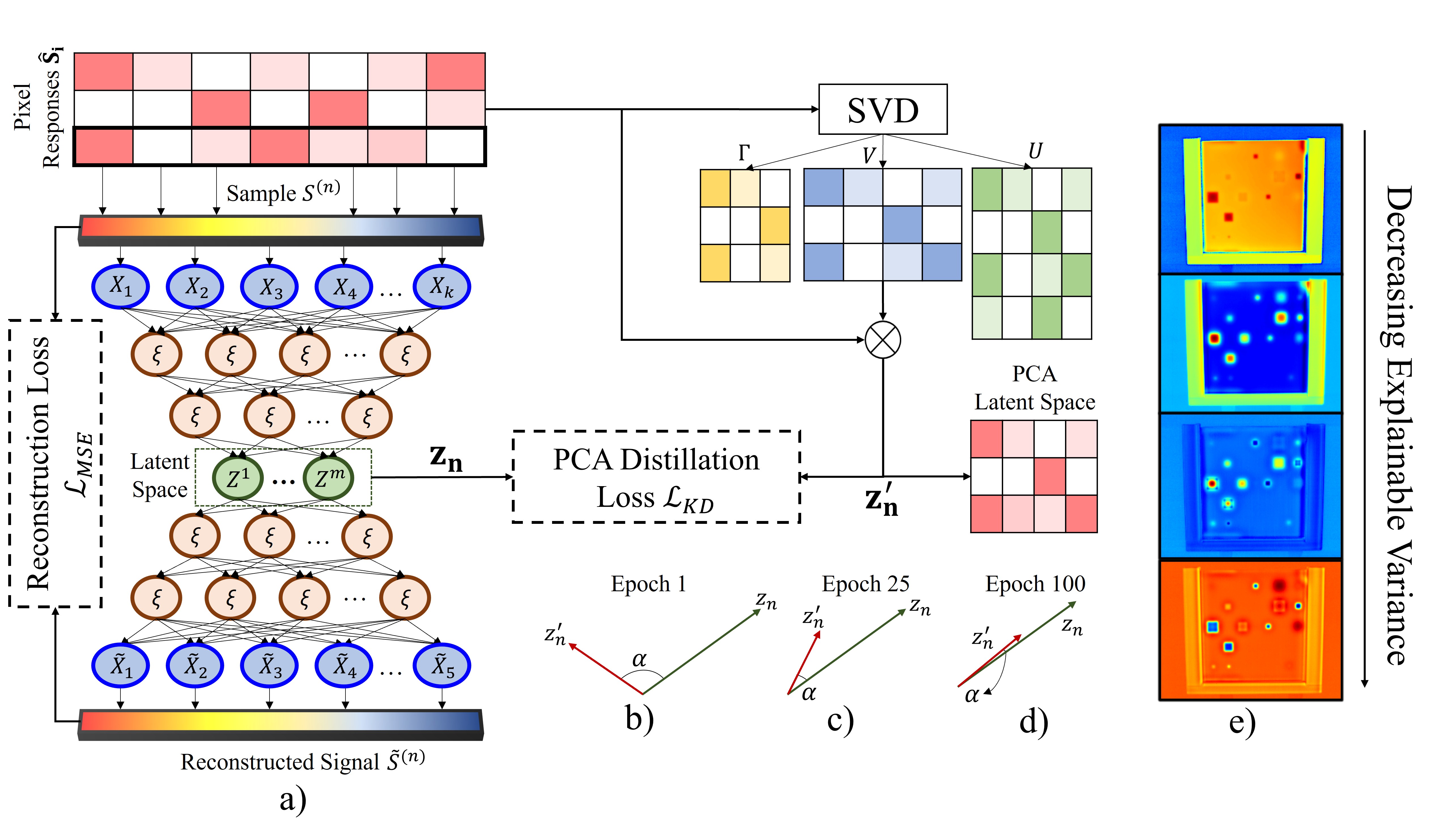}
    \caption{Overview of the proposed PCA-Guided Autoencoder (AE) framework. a) The autoencoder reconstructs pixel-wise temporal signals while learning a compressed latent representation $\mathbf{Z}_n$. During training, alignment between the AE latent space, $\mathbf{Z}_n$, and the PCA latent space $\mathbf{Z}^{'}_{n}$ is enforced progressively via PCA distillation loss, b)-d). e) The resulting latent representation captures the non-linearities in thermographic data with a structure defined by decreasing explainable variance.}
    \label{fig:framework}
\end{figure*}

\begin{equation}
    P_{k} = \mathbf{\hat{S}}v_{k},
\end{equation}

\noindent where $v_k$ denotes the $k^{th}$ column of $V$.

PCA falls under unsupervised machine learning models for extracting high-level features from the input AIRT data. However, its linear nature limits its ability to capture the complex, non-linear thermal patterns in AIRT sequences. Subtle, non-linear patterns are often caused by irregular heat diffusion around defective regions and  cannot be fully modeled by PCA linear projections. Thus, autoencoders, also known as AEs, emerged as a non-linear, learning-based alternative for dimensionality reduction and enhanced defect visualization. AEs are neural networks trained in an unsupervised manner to reconstruct their inputs through a low-dimensional latent representation. The network architecture of AEs generally comprises two key components, as shown in Fig. \ref{fig:ae_arch}. The AE encoder compresses the pixel responses, $S^{(n)}$, into a latent space of lower dimensionality defined by

\begin{equation}
    \mathbf{z}_n = f_{\boldsymbol{\theta}}(S^{(n)}),
\end{equation}

\noindent where $f_{\boldsymbol{\theta}}(\cdot)$ denotes the encoder function parameterized by weights $\theta$, and $\mathbf{z}_n \in \mathbb{R}^d$ represents the latent vector. On the other hand, the AE decoder reconstructs the input from the compressed latent vector $\mathbf{z}_n$ by

\begin{equation}
    \tilde{S}^{(n)} = g_{\boldsymbol{\phi}}(\mathbf{z}_n),
\end{equation}

\noindent where $g_{\boldsymbol{\phi}}(\cdot)$ is the decoder function with parameters $\phi$, and $\tilde{S}^{(n)}$ is the reconstructed signal. The latent space is used to formulate the latent images as a compressed representation for subsequent defect analysis. Various AEs have been proposed to tackle non-linear AIRT data, including 2D \cite{autoencoder} and 1D convolutional autoencoders \cite{1d_cnn}. Nevertheless, standard AEs often produce unstructured latent spaces, which can hinder their effectiveness in downstream AI-driven defect analysis tasks.

\section{Methodology} \label{sec:methodology}
\subsection{PCA-Guided Autoencoding}
In a standard AE, input pixel responses of an inspection sequence are passed to the network, which is trained in a self-supervised manner to reconstruct the original signal. However, multiple latent spaces can yield similar reconstructions, leading to inconsistencies in the learned latent vectors. To illustrate, Fig. \ref{fig:latents} shows the images defined using the first component of a conventional AE latent space, where it can be observed that the resulting representations vary across training runs, highlighting the need for structured latent space learning.

The proposed framework mitigates the aforementioned challenge and is outlined in Fig. \ref{fig:framework}. PCA principal components hold two important properties that define the structural consistency of its principal components: 1- The principal components tend to be orthogonal and uncorrelated, and 2- they are ordered in directions of descending explainable variance. For instance, the first principal component maximizes the variance, the second principal component the second most, and so on. Accordingly, the proposed PCA-Guided autoencoding aims to simultaneously capture the structural properties of the principal components and harness the non-linear nature of AEs for enhanced defect visibility. To do so, the conventional training objective of AEs is modified such that the PCA-Guided AE not only learns to reconstruct the input, $S^{n}$, from a latent space, $\mathbf{z}_{n}$, but also enforces alignment between $\mathbf{z}_{n}$ and the PCA latent vector, $\mathbf{z}^{'}_{n}$, by knowledge distillation. This proposed training objective is defined as

\begin{equation}
\boldsymbol{\theta}^*,\,\boldsymbol{\phi}^*
= \arg\min_{\boldsymbol{\theta},\,\boldsymbol{\phi}}
\left( 
      \underbrace{\mathcal{L}_{\text{rec}}\bigl(\tilde{S}^{(n)},\,S^{(n)}\bigr)}_{\text{Reconstruction}}
      \;+\;
      \underbrace{\mathcal{L}_{\text{KD}}\bigl(\mathbf{z}_{n},\,\mathbf{z}'_{n}\bigr)}_{\substack{\text{Knowledge} \\ \text{Distillation}}}
\right),
\label{eq:obj}
\end{equation}

\begin{algorithm}[t]
\caption{PCA-Guided Autoencoder Training}
\begin{algorithmic}[1]
\STATE \textbf{Input:} Standardized pixel responses matrix $\mathbf{\hat{S}}$
\STATE \textbf{Outputs:} Trained encoder $\theta^*$ \\
\hspace{1.35cm} Trained decoder $\phi^*$ \\
\hspace{1.35cm} Trained AE latent space $\mathbf{z}_{n}$
\STATE \textbf{Initialize:} Encoder $f_\theta$ and decoder $g_\phi$
\STATE Apply SVD on $\mathbf{\hat{S}}$ to obtain $\mathbf{z}'_n$
\WHILE{not converged}
    \FOR{Sample batch $S^n$ \textbf{in} $\mathbf{\hat{S}}$}
        \STATE \textbf{Encode} $\mathbf{z}_n \leftarrow f_{\boldsymbol{\theta}}(S^{(n)})$
        \STATE \textbf{Decode} $\tilde{S}^{(n)} \leftarrow g_{\boldsymbol{\phi}}(\mathbf{z}_n)$
        \STATE \textbf{Evaluate} Reconstruction loss $\mathcal{L}_{\text{rec}}(S^{(n)}, \tilde{S}^{(n)})$
        \STATE \textbf{Evaluate} Knowledge distillation loss $\mathcal{L}_{\text{KD}}(\mathbf{z}_{n},\,\mathbf{z}'_{n})$
        \STATE \textbf{Find} Total loss $\mathcal{L} = \mathcal{L}_{\text{rec}} + \mathcal{L}_{\text{KD}}$
        \STATE \textbf{Update} $\theta^* \leftarrow \theta - \eta \nabla_\theta \mathcal{L}, \quad
                      \phi^* \leftarrow \phi - \eta \nabla_\phi \mathcal{L}$
    \ENDFOR
\ENDWHILE
\STATE \textbf{return} $\theta^*$, $\phi^*$, $\mathbf{z}_{n}$
\end{algorithmic}
\label{alg:pca_guided_ae}
\end{algorithm}

\begin{figure*}[!ht]
    \centering
    \includegraphics[keepaspectratio=true,scale=0.45, width=0.9\linewidth]{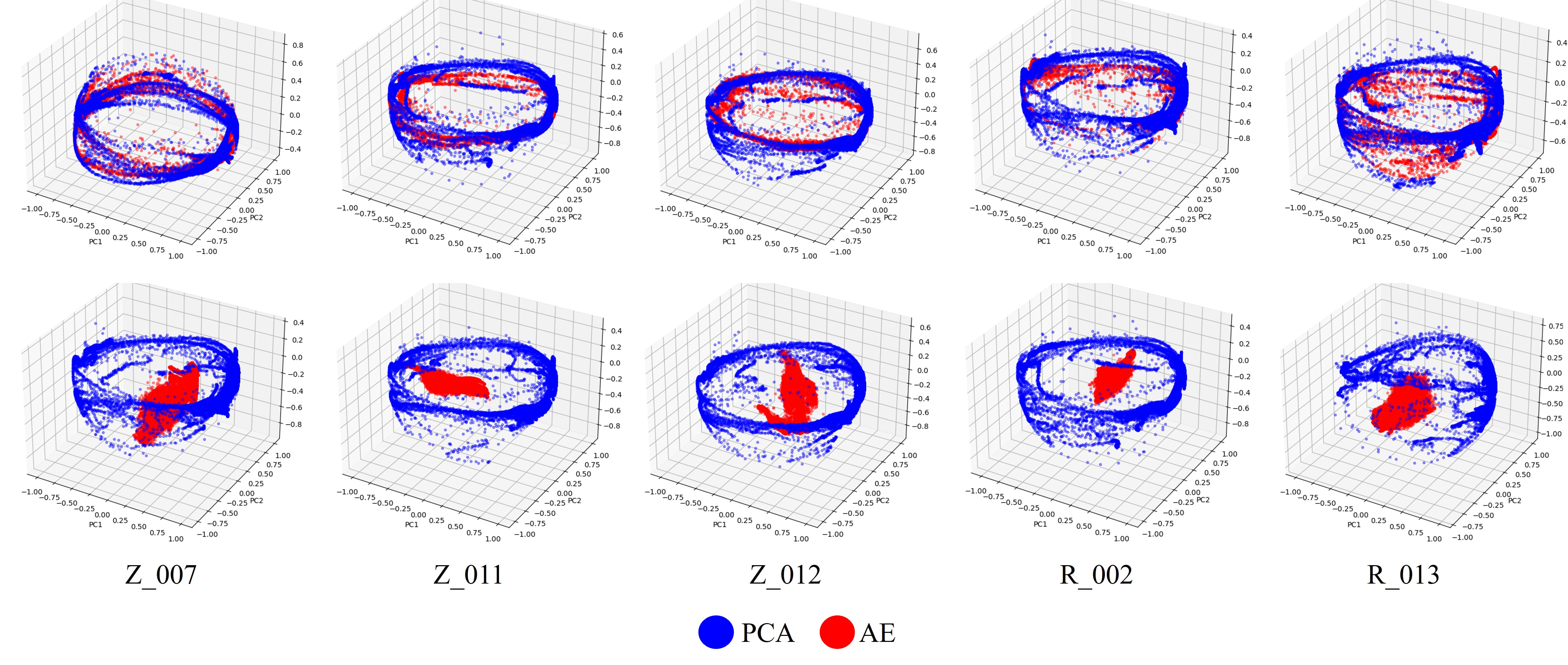}
    \caption{Normalized latent space of the PCA-Guided AE (top row) and standard AE (bottom row) compared to PCA principal components. Unlike the latent space of the standard AE, the latent vector of the PCA-Guided AE maintains consistency across samples and is aligned with the principal components.}
    \label{fig:alignment_samples}
\end{figure*}

\noindent where $\theta^*$ and $\phi^*$ are the optimal encoder and decoder parameters, respectively. The dual-objective formulation enables the AE to capture non-linear thermal patterns in input samples while constructing a learned latent space that possesses the structural properties of PCA orthogonal functions.

Having formulated the desired training objective, the training procedure of the AE is outlined in Algorithm \ref{alg:pca_guided_ae}. Given the matrix of standardized pixel responses, $\mathbf{\hat{S}}$, SVD is applied for the entire inspection sequence to obtain its compressed PCA latent representation, $\mathbf{z}^{'}_{n}$. $\mathbf{z}^{'}_{n}$ serves as a structural reference for guiding the AE’s latent space. Each training iteration samples a batch of pixel-wise temporal signals, $S^{(n)}$, from $\mathbf{\hat{S}}$, which are passed through the encoder to obtain the compressed representation $\mathbf{z}_n$. The decoder reconstructs the original signals from $\mathbf{z}_n$, while the encoder is simultaneously guided to align its outputs with $\mathbf{z}'_n$. During each training step, the network performs a forward pass for each sampled input, computes the reconstruction and alignment objectives, and updates the encoder and decoder parameters via backpropagation. The reconstruction component ensures that the learned latent features preserve non-linearities for signal fidelity, while the alignment term promotes structural consistency, see Fig. \ref{fig:framework}b-d. Both the encoder and decoder are composed of fully connected layers with ReLU activations. After training, the structured latent space, $\mathbf{z}_{n}$, is used to construct the latent images for enhanced visualization and subsequent defect analysis. It is worth highlighting that the proposed framework is agnostic to the network architecture. Meaning, the fully connected layers can be replaced by convolutional layers. In this work, however, a fully connected architecture is adopted due to its simplicity for analyzing pixel-wise temporal signals. The next section discusses the PCA distillation loss to fully formulate the training loss function that meets the objective defined in Eq. \ref{eq:obj}.

\subsection{Proposed Loss Function}
The proposed loss function plays the most important role in training the PCA-Guided AE. The formulated objective in Eq. \ref{eq:obj} highlights that a joint optimization of the network parameters involves achieving the objective of reconstruction and alignment via knowledge distillation. Several knowledge distillation losses are studied in the literature, including Kullback–Leibler (KL) divergence and mean squared error (MSE). However, those distillation losses either over-penalise magnitude differences (MSE) or assume a probabilistic output (KL-divergence), which is ill-suited to the desired, deterministic latent vectors. Instead, the PCA distillation loss is introduced, which is defined by the cosine similarity and is formulated by

\begin{equation}
\mathcal{L}_{\text{KD}} =
1 -
\frac{
\langle \mathbf{z}_n,\, \mathbf{z}'_n \rangle
}{
\|\mathbf{z}_n\|_2\, \|\mathbf{z}'_n\|_2
},
\label{eq:kd_loss}
\end{equation}

\noindent where $\langle \cdot,\cdot \rangle$ is the dot product. The intuition behind the PCA distillation loss is that angular deviation, $\alpha$, is penalized and directional alignment between $\mathbf{z}_n$ and $\mathbf{z}'_n$ is enforced. Fig. \ref{fig:samples} compares the normalized latent vectors of the PCA-guided AE and standard AE to the principal components, on samples from the IRT-PVC dataset \cite{irt_depth}. Notice that the latent space of the PCA-Guided AE maintains consistent structure and alignment with the PCA principal components across the inspected samples. Besides, the PCA distillation loss offers two practical advantages: 1- scale invariance, since the loss operates on normalized latent vectors, and 2- structural adherence, as it preserves the orthogonality encoded by the principal components of PCA.

On the other hand, the reconstruction loss, $\mathcal{L}_{\text{rec}}$, is the mean squared error (MSE), which measures the closeness of the reconstructed signal $\tilde{S}^{(n)}$ to the original input $S^{(n)}$. The reconstruction loss guides the network to reconstruct the original signal from the compressed latent vector to capture the non-linear patterns in the input thermographic signals. $\mathcal{L}_{\text{rec}}$ is formulated as

\begin{equation}
\mathcal{L}_{\text{rec}} =
\frac{1}{N} \sum_{i=1}^{N} \| \tilde{S}^{(n)} - S^{(n)} \|_2^2,
\label{eq:recon_loss}
\end{equation}

\noindent where $N$ is the number of pixels in the batch and $S_i^n$ denotes the temporal response of the $i^{th}$ pixel. Combining both objectives, the total loss used to train the PCA-Guided AE is expressed as

\begin{equation}
\mathcal{L}_{\text{total}} =
\mathcal{L}_{\text{rec}} +
\alpha\, \mathcal{L}_{\text{KD}},
\label{eq:total_loss}
\end{equation}

\noindent where $\alpha=1.0$ is a hyperparameter that balances the influence of the distillation loss relative to the reconstruction objective.

Given the PCA-Guided AE training procedure and its loss function, a Bayesian optimization process is applied to the AE to select its hyperparameters, which include the number of layers, latent vector size, number of epochs, batch size, and learning rate. The obtained optimal AE hyperparameters suggest a latent vector size of 64, three encoding, and decoding layers. In addition, the network is also trained using ADAM optimizer, with a learning rate of $1e^{-4}$, and a batch size of 512 on each inspection sequence.

\section{Experiments} \label{sec:exps}
\subsection{Experimental Setup}

\begin{figure}[t]
    \centering
    \includegraphics[keepaspectratio=true,scale=0.45, width=0.7\linewidth]{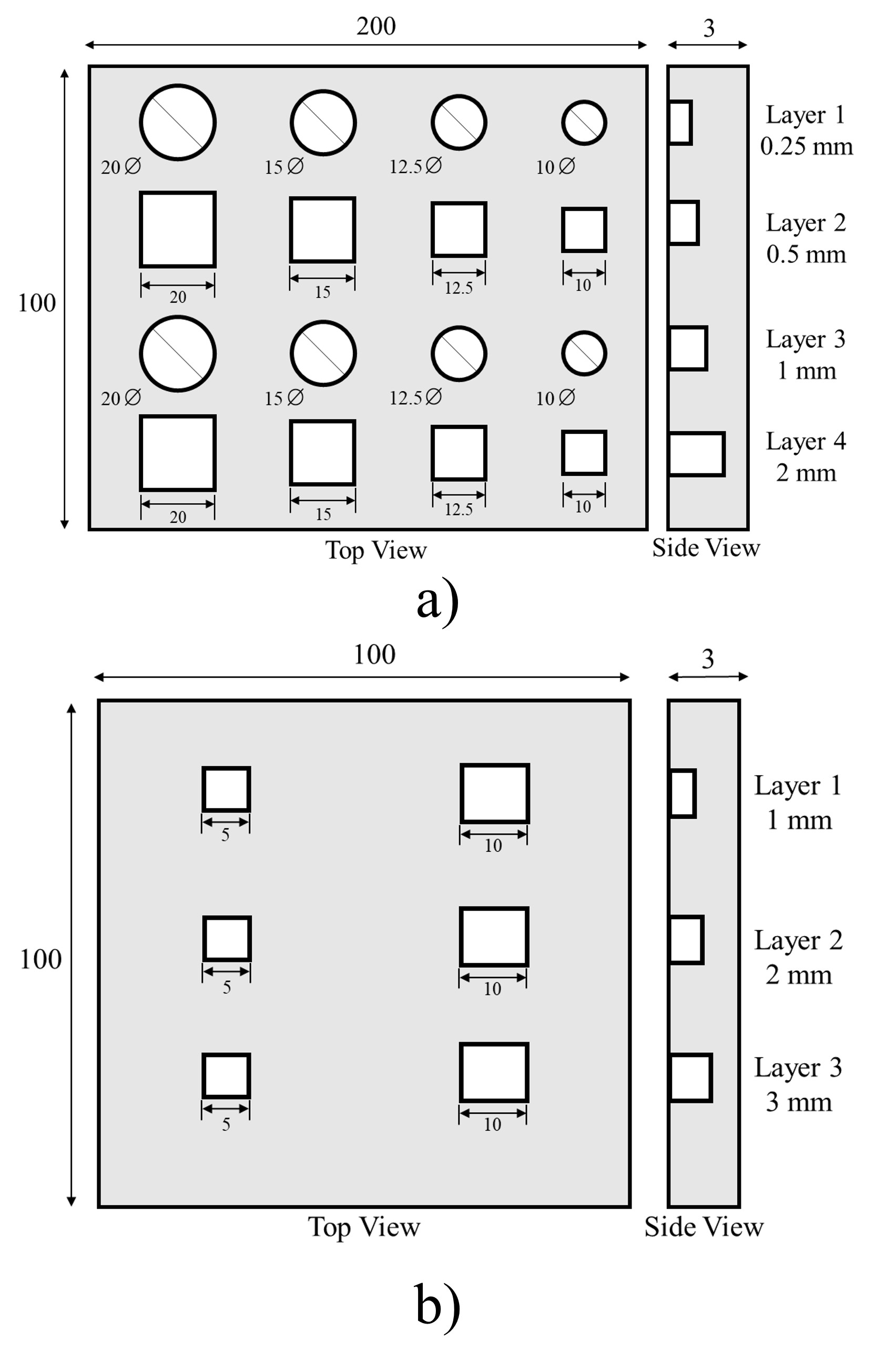}
    \caption{Validation of PCA-Guided AE on CFRP and PLA samples shown in a) and a PLA sample shown in b). The dimensions are expressed in mm.}
    \label{fig:samples}
\end{figure}

\begin{figure}[b]
    \centering
    \includegraphics[keepaspectratio=true,scale=0.45, width=\linewidth]{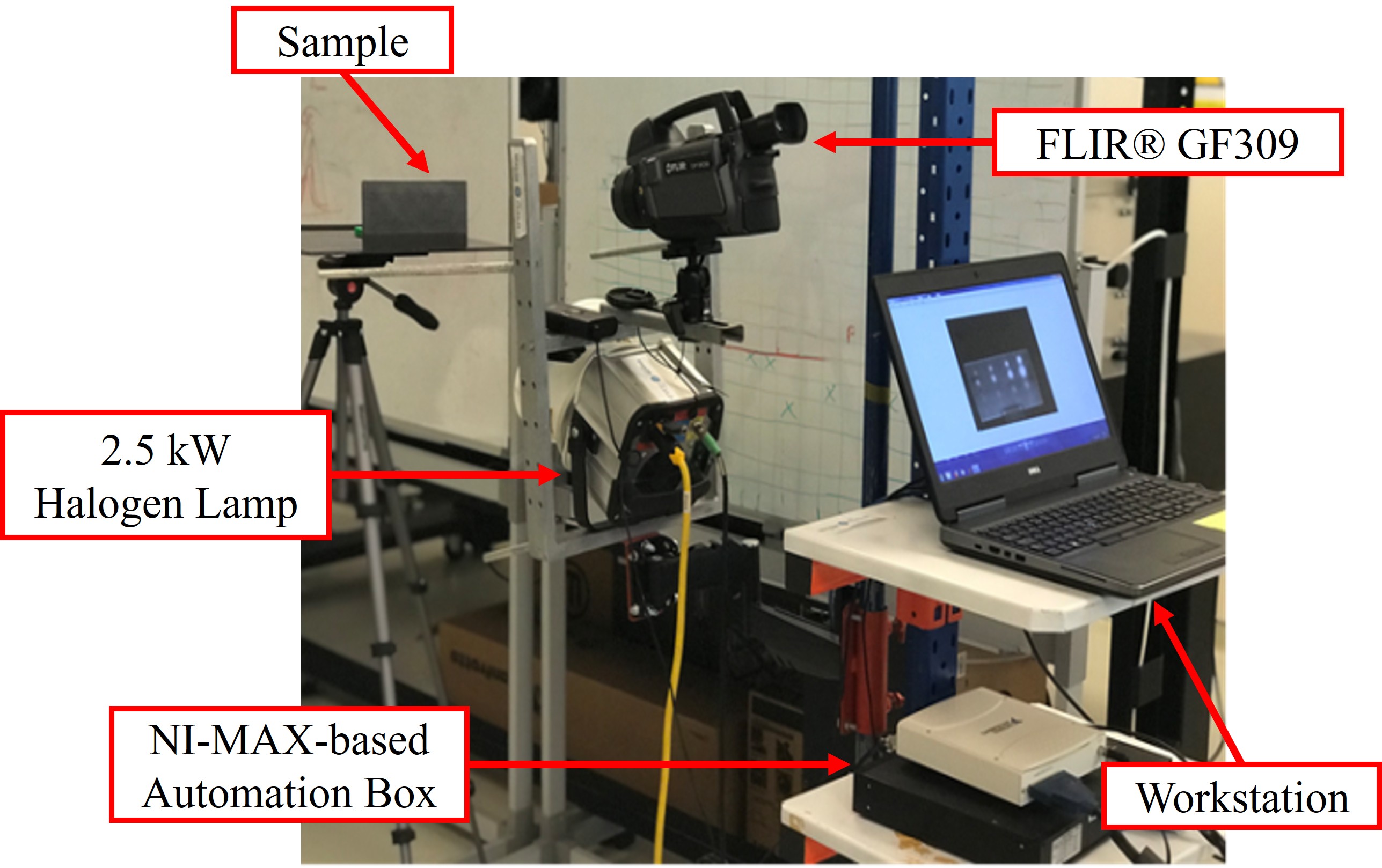}
    \caption{Data collection setup for evaluating the PCA-Guided autoencoding framework.}
    \label{fig:setup}
\end{figure}

\begin{table*}[p]
\centering
\renewcommand{\arraystretch}{2}
\caption{Qualitative comparisons between state-of-the-art AIRT dimensionality reduction techniques: TSR, PCA, DAT \cite{dat}, and 1D-DCAE-AIRT \cite{1d_cnn}, against the proposed PCA-Guided AE.}
\resizebox{\textwidth}{!}{%
\begin{tabular}{|>{\centering\arraybackslash}m{2.5cm}|>{\centering\arraybackslash}m{4cm}|>{\centering\arraybackslash}m{4cm}|>{\centering\arraybackslash}m{4cm}|}
\hline
\textbf{Method} & \textbf{CFRP} & \textbf{PLA} & \textbf{PVC} \\ \hline

\textbf{Raw} &
{\vspace{3pt}\includegraphics[width=3.75cm, height=2.5cm]{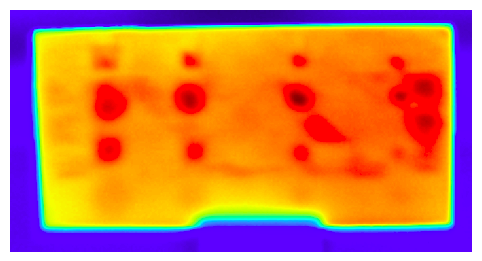}} &
{\vspace{3pt}\includegraphics[width=3.75cm, height=2.5cm]{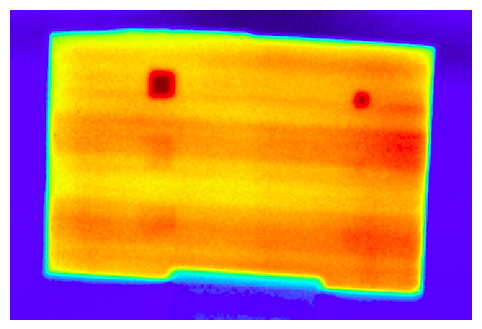}} &
{\vspace{3pt}\includegraphics[width=3.25cm, height=2.5cm]{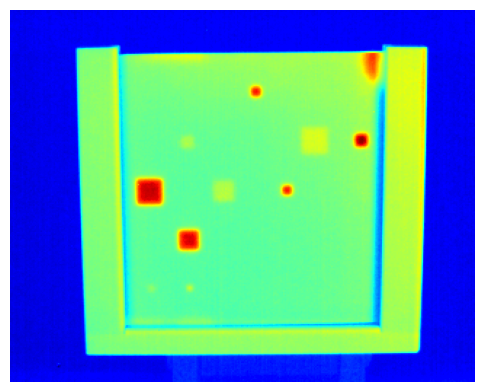}} \\ \hline

\textbf{TSR} &
{\vspace{3pt}\includegraphics[width=3.75cm, height=2.5cm]{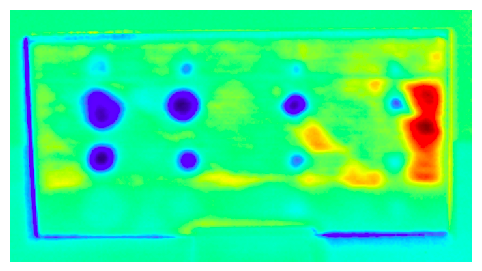}} &
{\vspace{3pt}\includegraphics[width=3.75cm, height=2.5cm]{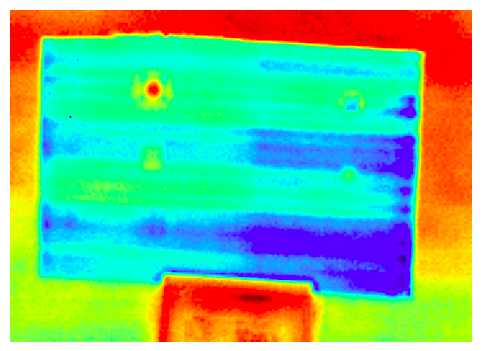}} &
{\vspace{3pt}\includegraphics[width=3.25cm, height=2.5cm]{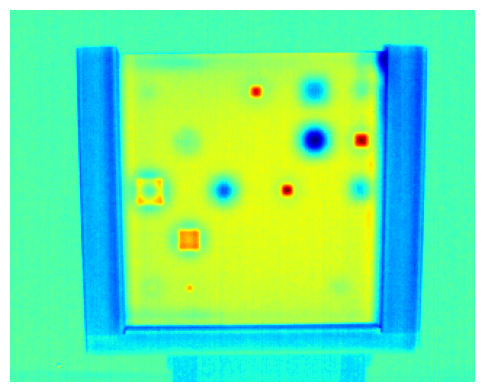}} \\ \hline

\textbf{PCA} &
{\vspace{3pt}\includegraphics[width=3.75cm, height=2.5cm]{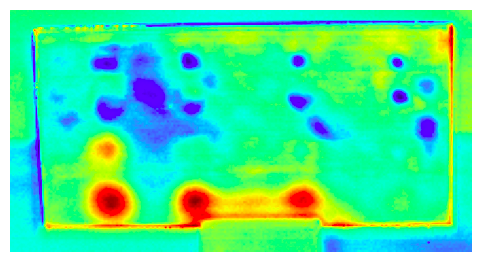}} &
{\vspace{3pt}\includegraphics[width=3.75cm, height=2.5cm]{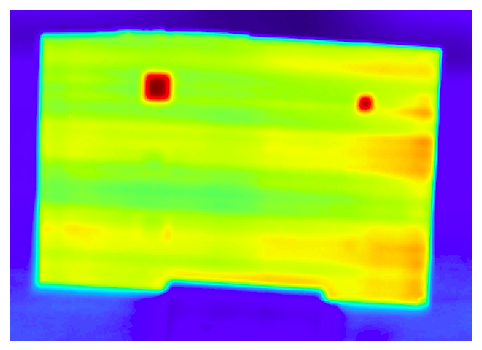}} &
{\vspace{3pt}\includegraphics[width=3.25cm, height=2.5cm]{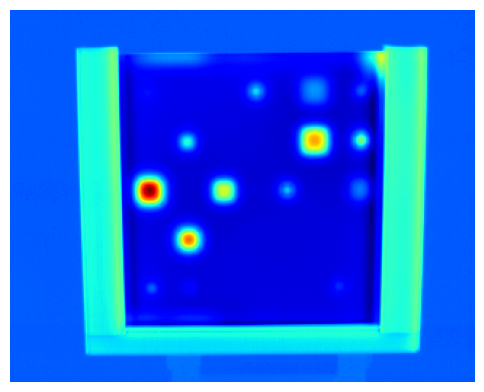}} \\ \hline

\textbf{DAT} &
{\vspace{3pt}\includegraphics[width=3.75cm, height=2.5cm]{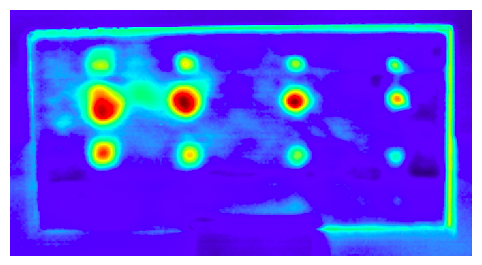}} &
{\vspace{3pt}\includegraphics[width=3.75cm, height=2.5cm]{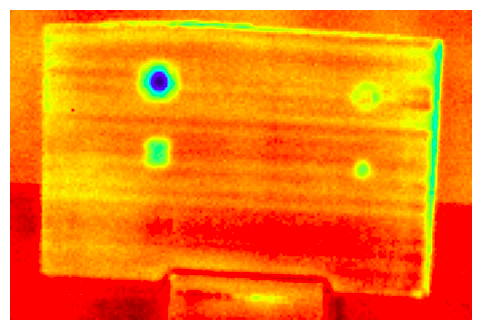}} &
{\vspace{3pt}\includegraphics[width=3.25cm, height=2.5cm]{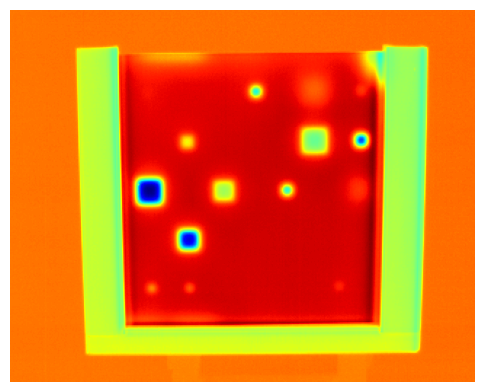}} \\ \hline

\textbf{1D-DCAE-AIRT} &
{\vspace{3pt}\includegraphics[width=3.75cm, height=2.5cm]{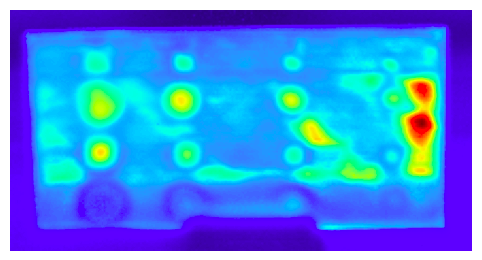}} &
{\vspace{3pt}\includegraphics[width=3.75cm, height=2.5cm]{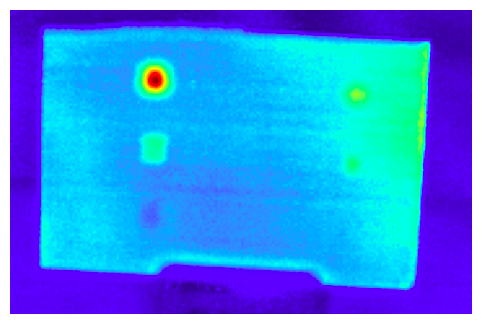}} &
{\vspace{3pt}\includegraphics[width=3.25cm, height=2.5cm]{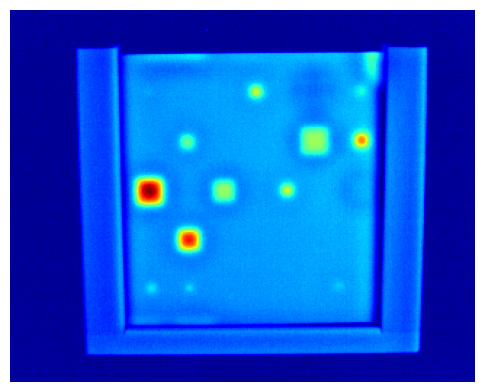}} \\ \hline

\textbf{Ours} &
{\vspace{3pt}\includegraphics[width=3.75cm, height=2.5cm]{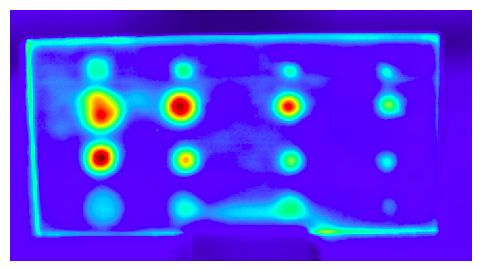}} &
{\vspace{3pt}\includegraphics[width=3.75cm, height=2.5cm]{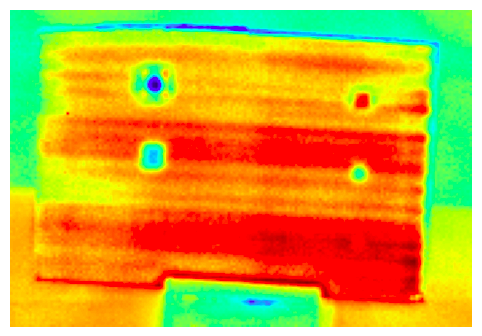}} &
{\vspace{3pt}\includegraphics[width=3.25cm, height=2.5cm]{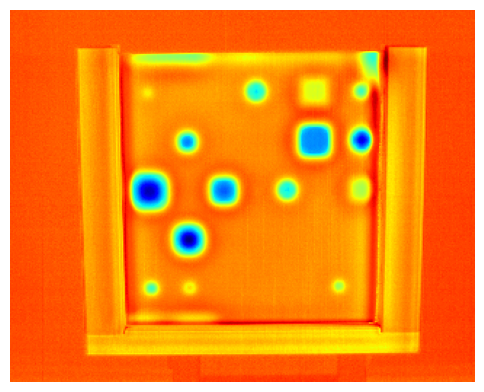}} \\ \hline

\end{tabular}
}
\label{table:qual_comparison}
\end{table*}

The proposed framework is evaluated on three samples: one CFRP and two PLA. 3D printing was utilized to produce the samples of the two low-conductivity materials. The samples are designed such that they simulate real-world defects and incorporate embedded flaws similar to those created using Teflon insertions discussed in \cite{yusra_4}. Samples 1 and 2 are generated with 16 artificial defects of different depths, sizes, and shapes. Similarly, sample 3 is generated with 6 square shaped defects at different depths, see Fig. \ref{fig:samples}. The defects are distributed at four depth layers, i.e. 0.25, 0.50, 1.0, 2.0 mm. The AIRT inspection sequences are generated using the setup shown in Fig. \ref{fig:setup}. The samples are excited by a heat pulse using a 230V, 2.5 kW halogen lamp controlled by an NI-MAX-based automation box as a signal generator for precise sequencing and control of the heat pulses. In addition, different pulse durations, 5, 10, and 15 seconds, where in total 9 inspection sequences are collected. GF309 from FLIR® featuring an Indium Antimonide InSb cooled thermal is utilized to acquire the thermal image sequences. The thermal sensor operates at a frequency of 25 Hz within the 3-5 $\mu$m infrared range and achieves a Noise Equivalent Temperature Difference (NETD) of 0.025K. The IR camera and the halogen lamp were positioned at 10 cm and 8 cm, respectively. In addition, the halogen lamp is placed at 15$\circ$ incidence angle. This configuration produces a spatial resolution of $0.4 \times 0.4$ mm/pixel while intentionally introducing non‑uniform heating to assess the robustness of the proposed PCA‑guided autoencoder.

In addition to the CFRP and PLA sequences, the PCA-Guided AE is also evaluated on the IRT-PVC dataset \cite{irt_depth}. The dataset comprises of 38 inspection sequences of 3D-printed PVC samples with back-drilled holes at depths ranging between 0.0 and 2.5 mm. Given the CFRP, PLA, and PVC inspection sequences, the experimental evaluation routine is two-fold. First, defect visibility and signal enhancement evaluation is conducted and presented in Section \ref{subsection:enhancement}. For this evaluation routine, the utilized metrics are contrast and SNR defined by

\begin{equation}
    \text{Contrast} = \frac{\left| \left(\frac{\sum_{i=1}^n X(i)_D}{n}\right) - \left(\frac{\sum_{j=1}^m X(j)_I}{m}\right) \right|}{\left(\frac{\sum_{i=1}^n X(i)_D}{n}\right) + \left(\frac{\sum_{j=1}^m X(j)_I}{m}\right)},
\end{equation}

\begin{equation}
    \text{SNR} = \frac{\left| \left(\frac{\sum_{i=1}^n X(i)_D}{n}\right) - \left(\frac{\sum_{j=1}^m X(j)_I}{m}\right) \right|}{\sigma_I},
\end{equation}

\begin{table*}[!ht]
\centering
\caption{Quantified contrast and SNR for the PCA-Guided AE benchmarked against state-of-the-art thermography dimensionality reduction methods, TSR, PCA, DAT \cite{dat}, and 1D-DCAE-AIRT \cite{1d_cnn}.}
\label{table:quantify_snr}
\resizebox{0.8\textwidth}{!}{%
\begin{tabular}{|c|c|c|c|c|c|c|c|}
\hline
\textbf{Sample} & \textbf{Metric} & \textbf{Raw} & \textbf{TSR} & \textbf{PCA} & \textbf{DAT} & \textbf{1D-DCAE-AIRT} & \textbf{Ours} \\ \hline
\multirow{2}{*}{CFRP} & Contrast & 0.221 & 0.305 & 0.311 & 0.366 & 0.391 & \textbf{0.497} \\ \cline{2-8} 
                            & SNR (dB) & 22.83 & 25.93 & 30.32 & 32.44 & 32.71 & \textbf{37.84} \\ \hline
\multirow{2}{*}{PLA} & Contrast & 0.296 & 0.205 & 0.281 & 0.426 & 0.431 & \textbf{0.483} \\ \cline{2-8} 
                            & SNR (dB) & 25.78 & 21.29 & 25.67 & 26.70 & 28.54 & \textbf{34.91} \\ \hline
\multirow{2}{*}{PVC} & Contrast & 0.324 & 0.505 & 0.381 & 0.526 & 0.531 & \textbf{0.588} \\ \cline{2-8} 
                            & SNR (dB) & 27.05 & 24.57 & 26.06 & 31.43 & 35.75 & \textbf{39.99} \\ \hline
\end{tabular}%
}
\end{table*}

\begin{table*}[!ht]
\centering
\caption{PCA-Guided AE validated and benchmarked using the proposed neural network-based evaluation, where the PCA-Guided AE latent representation is utilized as input to a segmentation neural network and the IoU serves as an evaluation metric.}
\label{tab:iou}
\resizebox{0.8\textwidth}{!}{%
\begin{tabular}{|c|c|c|c|c|c|c|}
\hline
\textbf{Method} & \textbf{Raw} & \textbf{TSR} & \textbf{PCA} & \textbf{DAT} & \textbf{1D-DCAE-AIRT} & \textbf{Ours}  \\ \hline
Validation IoU  & 0.751        & 0.738        & 0.743        & 0.209        & 0.763                 & \textbf{0.780} \\ \hline
Testing IoU     & 0.759        & 0.734        & 0.721        & 0.221        & 0.754                 & \textbf{0.784} \\ \hline
\end{tabular}%
}
\end{table*}

\noindent where $n$ is the number of pixels in the defective region $X_{D}$ and $X(i)_{D}$ is the $i^{th}$ pixel value in $X_{D}$. Similarly, $m$  is the number of pixels in the sound regions $S_{I}$ and $X(j)_{I}$ is the $j^{th}$ pixel value in $X_{I}$. $\sigma_{I}$ represents the standard deviation of the pixel region in the sound area $X(j)_{I}$. The second evaluation, discussed in Section \ref{subsection:nn_eval}, employs the proposed neural network-based evaluation, where the PCA-Guided latent space is utilized as a representation for a segmentation U-Net and the evaluation metric is the intersection over union (IoU) defined as 

\begin{equation}
    \text{IoU} = \frac{|P \cap G|}{|P \cup G|},
\end{equation}

\noindent where $P$ and $G$ are the predicted and ground truth segmentation masks. This evaluation assesses the effectiveness of the proposed PCA-Guided AE latent representation for AI-based defect analysis and highlights the importance of constructing structured latent spaces. Note that this evaluation is conducted on the IRT-PVC dataset since it was designed for training deep neural networks. In addition, both evaluation routines present benchmarks of the PCA-Guided AE against state-of-the-art AIRT dimensionality reduction methods, including TSR, PCA, DAT \cite{dat}, and 1D-DCAE-AIRT \cite{1d_cnn}.

\subsection{Signal Enhancement Evaluation} \label{subsection:enhancement}
AIRT dimensionality reduction methods serve the purpose of reducing data dimensionality and enhancing defect visualization. Generally, enhanced defect visibility is correlated with improved defect analysis. The presented signal enhancement evaluation demonstrates the effectiveness of the PCA-Guided AE in enhancing defect visibility. For instance, Table \ref{table:qual_comparison} qualitatively compares the PCA-Guided latent representation to the compressed representations of PCA, TSR, DAT \cite{dat}, and 1D-DCAE-AIRT \cite{1d_cnn} on CFRP, PLA, and PVC samples. Note that DAT and 1D-AIRT-CAE serve as standard AIRT AEs, where the former is composed of fully connected layers and the latter follows a convolutional architecture. The visualizations show that raw thermograms show some but not all defects. For instance, the raw thermogram from the CFRP inspection sequence only shows 11 visible defects out of 16. The TSR representation falls short with only 9 visible defects. On the other hand, PCA, DAT, and 1D-DCAE-AIRT enhance defect clarity compared to TSR; however, the last row of defects is merely visible. In contrast, all of the 16 defects in the CFRP sample are clearly distinguishable in the PCA-Guided AE latent images. This analysis is also applicable to the PLA and PVC specimens, where some defects are not highlighted using the previous approaches.

To quantify signal enhancement provided by the PCA-Guided AE, Table \ref{table:quantify_snr} provides the mean contrast and SNR of the proposed latent representation along with state-of-the-art AIRT representations as a benchmark on all CFRP, PLA, and PVC inspection sequences. Several findings can be inferred from the results. First, the obtained results demonstrate that the PCA-Guided AE provides state-of-the-art results in terms of contrast and SNR. Second, the PCA-Guided AE tends to capture the non-linear patterns in thermographic sequences, even though the latent space tends to be aligned with PCA principal components. Finally, the PCA-Guided AE outperforms DAT, a model with a similar architectural design but without latent space alignment. This highlights that guiding the latent space toward PCA principal components not only promotes structural organization but also contributes to improved defect visibility.

\begin{figure}[t]
    \centering
    \includegraphics[keepaspectratio=true,scale=0.45, width=\linewidth]{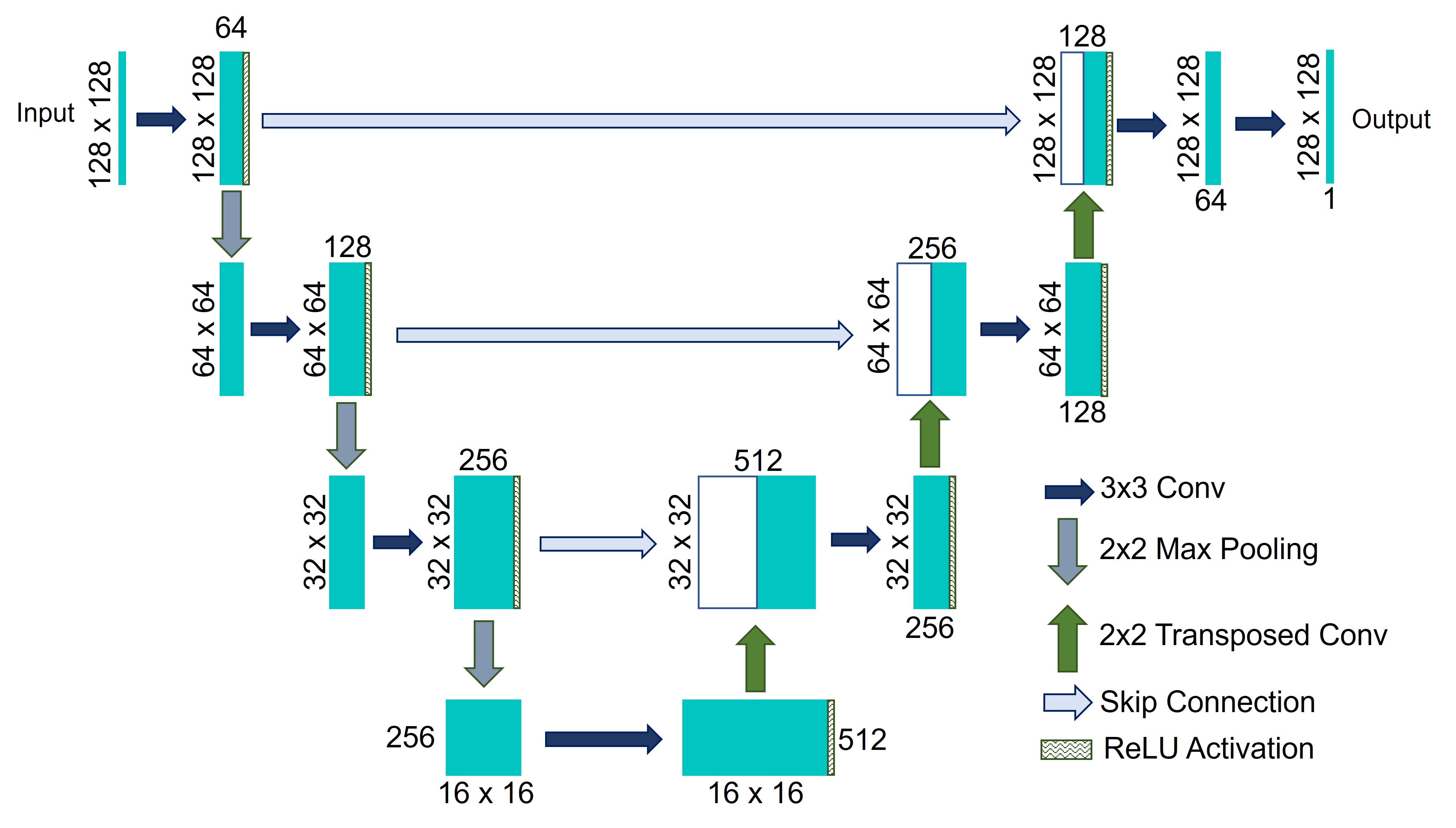}
    \caption{U-Net segmentation network used for the neural network-based evaluation.}
    \label{fig:unet}
\end{figure}

\subsection{Neural Network-Based Evaluation} \label{subsection:nn_eval}
The contrast and SNR serve as metrics to quantify signal enhancement and defect clarity of AIRT data compression methods. However, as AI methodologies are currently adopted in AIRT, it is essential to test the effectiveness of AIRT compressed representations as inputs to neural networks. Thus, the second part of the evaluation involves the proposed neural network-based evaluation, where the PCA-Guided AE latent images are utilized as an input to a U-Net segmentation neural network, shown in Fig. \ref{fig:unet}. The network mimics the U-Net tested in \cite{pt_dataset} and the IoU serves as the evaluation metric. The IRT-PVC dataset follows the same split reported in \cite{irt_depth}, where it is divided to 26 training, 6 validation, and 6 testing sequences. Note that improved network performance and higher IoUs correlate to a representation with enhanced defect clarity and, more importantly, a structurally consistent latent space.

Table \ref{tab:iou} provides the validation and testing IoUs of the U-Net obtained when the PCA-Guided AE acts as input to the U-Net. Table \ref{tab:iou} also presents the validation and testing IoUs for the previously tested AIRT representations as inputs, to provide benchmarks for the proposed framework. The results show that the PCA-Guided AE outperforms classical AIRT dimensionality reduction methods, i.e. PCA and TSR, with an increased IoU of 5\%. This highlights the ability of the proposed AE to capture non-linear thermal patterns for enhanced defect analysis. Compared to learning-based approaches, i.e. DAT and 1D-DCAE-AIRT, the U-Net provides an increased IoU when relying on the PCA-Guided AE representation as an input. Interestingly, the U-Net achieves an IoU of approximately 0.20 when DAT representation is used as an input. Note that the DAT representation resembles the standard AIRT AE counterpart of the proposed PCA-Guided AE, where unstructured latent spaces are constructed. This motivates the need for structured learning and autoencoding of thermographic sequences, achieved by the introduced PCA-Guided autoencoding framework. Finally, the PCA-Guided AE outperfoms the 1D-DCAE-AIRT with 2\% improvement in IoU. Even though the network performance is improved by a small margin, the network architecture of the PCA-Guided AE is simpler and faster to train compared to the 1D-DCAE-AIRT. To illustrate, training the PCA-Guided AE using RTX 3060 Laptop GPU on $\textit{R\_006}$ sequence from the IRT-PVC dataset takes 42.33 seconds, while the 1D-DCAE-AIRT required 618.67 seconds to train. This shows that training time is reduced by a factor of $14\times$, while providing an improved representation for neural networks with enhanced defect clarity.

\section{Conclusions} \label{sec:conc}
Active Infrared thermography (AIRT) emerged as a powerful tool in NDT for detecting hidden defects in industrial components. A challenge in AIRT is the large dimensionality of thermographic data. Classical AIRT dimensionality reduction methods, such as PCA and TSR, struggle in capturing the complex and intricate patterns in thermographic sequences. On the other hand, learning-based AIRT autoencoders (AEs) construct latent spaces that lack structure and consistency. These drawbacks limit the performance of AIRT AI-based characterization frameworks when relying on the aforementioned representations. Therefore, this paper presented PCA-Guided autoencoding for learning compact and structured latent spaces from AIRT sequences. The proposed work introduced a novel loss function, PCA distillation loss, that guides the latent space of non-linear AIRT AEs to align with the principal components derived from PCA, achieving structured and semantically meaningful embeddings. To validate the effectiveness of the PCA-Guided autoencoding framework, a neural network-based evaluation was introduced to test the effectiveness of the PCA-Guided AE representation. Extensive experiments on CFRP, PLA, and PVC samples demonstrated that PCA-Guided autoencoding achieves representations with enhanced contrast and SNR. More importantly, defect segmentation neural networks tend to witness improved performances when relying on the PCA-Guided AE latent representation as input compared to state-of-the-art thermographic representations. Future research efforts will be directed toward enhancing AIRT AE architectures and incorporating attention mechanisms to capture long-range dependencies in AIRT data. In addition, future work will investigate efficient training strategies for fast training of complex and deep AIRT AE architectures.

\bibliographystyle{IEEEtran}
\bibliography{main.bib}

\end{document}